\documentclass[sigconf]{acmart}
\renewcommand\footnotetextcopyrightpermission[1]{} 
\usepackage[ruled,vlined]{algorithm2e}
\AtBeginDocument{%
  \providecommand\BibTeX{{%
    \normalfont B\kern-0.5em{\scshape i\kern-0.25em b}\kern-0.8em\TeX}}}

\begin{document}
\fancyhead{}
\title{SPot: A tool for identifying operating segments \\in financial tables}


\author{Zhiqiang Ma}
\affiliation{%
  \institution{S\&P Global}
  \city{New York}
  \country{NY}}
\email{zhiqiang.ma}
\email{@spglobal.com}

\author{Steven Pomerville}
\affiliation{%
  \institution{S\&P Global}
  \city{New York}
  \country{NY}
}
\email{steven.pomerville}
\email{@spglobal.com}

\author{Mingyang Di}
\affiliation{%
  \institution{SiriusXM}
  \city{New York}
  \country{NY}
}
\email{kingsley.di}
\email{@siriusxm.com}

\author{Armineh Nourbakhsh}
\affiliation{%
  \institution{S\&P Global}
  \city{New York}
  \country{NY}
}
\email{armineh.nourbakhsh}
\email{@spglobal.com}

\renewcommand{\shortauthors}{Ma, et al.}

\begin{abstract}
 In this paper we present \textit{SPot}, an automated tool for detecting operating segments and their related performance indicators from earnings reports. Due to their company-specific nature, operating segments cannot be detected using taxonomy-based approaches. Instead, we train a Bidirectional RNN classifier that can distinguish between common metrics such as ``revenue'' and company-specific metrics that are likely to be operating segments, such as ``iPhone'' or ``cloud services''. SPot surfaces the results in an interactive web interface that allows users to trace and adjust performance metrics for each operating segment. This facilitates credit monitoring, enables them to perform competitive benchmarking more effectively, and can be used for trend analysis at company and sector levels.
\end{abstract}

\ccsdesc[500]{Information systems~Information retrieval}
\ccsdesc[300]{Applied computing~Document management and text processing}
\ccsdesc[300]{Information systems~Web applications}

\keywords{Financial Reports, SEC Reports, Earnings Reports, Operating Segments, Web Application, Products and Services, Business Divisions}


\maketitle

\section{Introduction}\label{intro}
Based on international financial reporting standards, an operating segment is a profit component of a business entity that has discrete financial information available and whose results are evaluated regularly by the entity's management for purposes of performance assessment\footnote{https://www.accountingtools.com/articles/2017/5/13/operating-segment}. A company's operating segments can be its products, services, business divisions, geographic locations, or assets (such as mines, reserves, wells, oilfields, etc.).\footnote{Not all assets qualify as operating segments. They can only be considered an operating segment if they function as an active source of  revenue.}

Credit analysts consistently monitor the performance of a business within each of its operating segments in order to determine major areas of risk or growth. For instance if a certain product is the main driver of profit for a given company, then a fall in net sales for that product might pose a financial risk to the company. Monitoring the performance of operating segments requires reading through lengthy financial reports and extracting each segment and its corresponding performance metric manually from tables. 

In this paper, we introduce \textbf{SPot} (or \textbf{S}\&\textbf{P} \textbf{O}perating segmen\textbf{T} extractor), a tool that ingests financial reports from public U.S. companies in real-time, processes each table, and identifies each row header or column header that is likely to express an operating segment. The corresponding rows/columns are then extracted, aggregated, and displayed to the end-user on an interactive UI that allows them to study, trace, and adjust the performance indicators associated with each operating segment.

At the most basic level, SPot's main task reduces to a binary classification problem at the table header level. Concretely, given a header, the system is supposed to identify whether it is likely to be associated with an operating segment, or a non-operating metric such as a financial metric, name of a board member, an office location, a debt schedule, etc. A few challenges complicate this task:
\begin{enumerate}
    \item Operating segments are company-specific, so a taxonomy-driven approach would not scale to unseen companies. 
    \item Named-entity recognition cannot be used because certain types of operating segments (such as business divisions or types of service) are not always named entities. On the other hand, actual named entities (such as the names of executive leaders) are often not operating segments.
    \item Positional cues and co-occurrence metrics fall short, because, even though operating segments tend to be expressed in the same tables, they are often co-located with non-operating items. For instance an Income Statement table might begin with operating segments and move on to standard financials such as Total Revenue and R\&D Expense.
\end{enumerate}

Table \ref{tab:examples} lists a few examples illustrating the above  challenges. To address these problems, we use a multi-stage process that filters tables down to those likely to include operating segments. Then each row is classified using a sequence model that utilizes selective masking in a way that minimizes overfitting.


\section{SYSTEM DESIGN}
\begin{figure}[]
  \centering
  \includegraphics[scale=0.3]{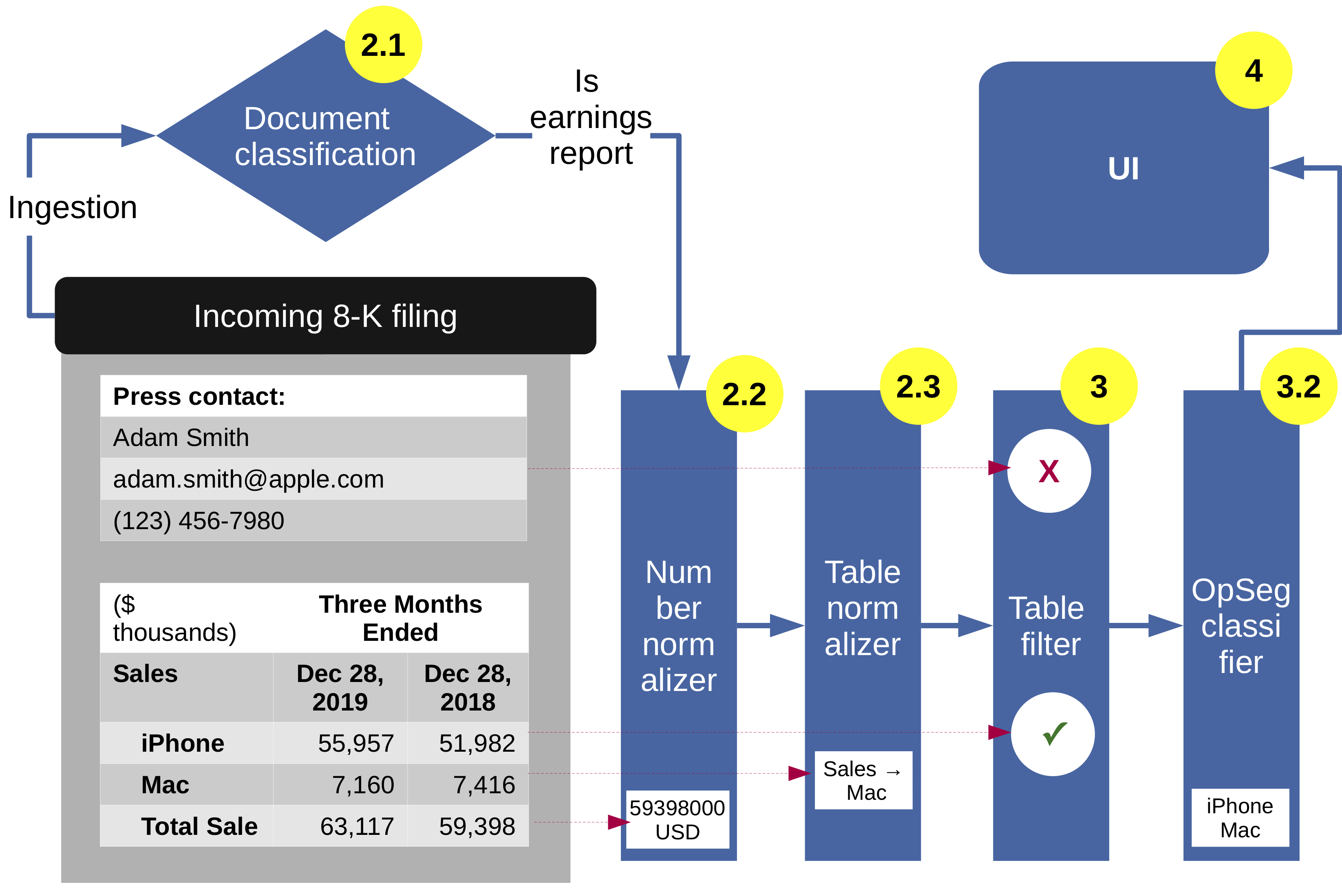}
  \caption{SPot's data-flow diagram. Solid blue arrows show the step by step process. Dashed red arrows show output examples for each step. Yellow circles include references to the section where each step is described.}
\label{fig:dataflow}
\end{figure}

\begin{figure*}[htp]
  \centering
  \includegraphics[scale=0.38]{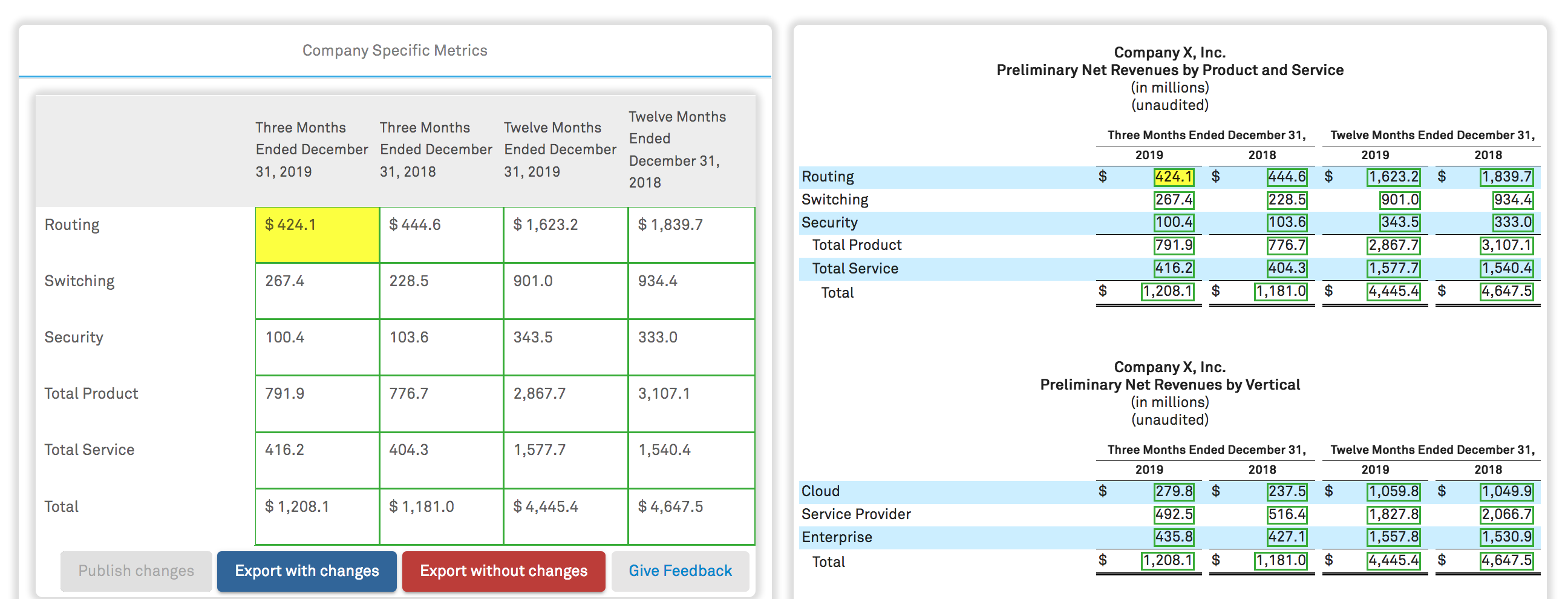}
  \caption{How the results are displayed to the user, linked to the original document, made adjustable and exportable. Clicking into any cell on the left-hand panel will cause it to be highlighted in yellow and the cursor on the right-hand panel jumps to the cell where the metric was extracted from (also highlighted in yellow). }
\label{fig:ui}
\end{figure*}

Figure \ref{fig:dataflow} demonstrates the data flow and the components in SPot, which the following subsections describe in more detail\footnote{Figures \ref{fig:structure}, \ref{fig:dataflow} and \ref{fig:ui} provide fabricated examples and do not reflect any company's actual financial reports.}.

\subsection{Ingestion and classification of documents}
SPot ingests earnings reports published by public U.S. companies to the SEC website\footnote{\url{https://www.sec.gov/cgi-bin/browse-edgar?action=getcurrent}}. As 8-K files are posted, they are ingested into the system through a proprietary sourcing service that uses SEC's RSS service. Unlike 10-Q and 10-K filings, 8-K filings are not limited to earnings reports, but can cover any material events that companies release to the public. A taxonomy-based classifier identifies 8-K filings that include earnings reports \cite{nourbakhsh2020}. 

\subsection{Normalization of numeric tables}
The reports are then processed through a normalization pipeline that performs the following steps: (1) Periods are identified and mapped to the company's fiscal calendar. For instance ``Three Months Ended March 30, 2020'' may be normalized to \texttt{``Q1 2020''}. (2) Financial numbers are identified and normalized to the scale expressed inside or in the vicinity of the table. For instance ``\$USD 14MM'' may be normalized to \texttt{``14,000,000.00 (USD)''}. (3) All other numbers are normalized to their raw form. For instance ``30 percent'' is normalized to \texttt{``30\%''}. A more detailed description of the ingestion and normalization pipeline is available in \cite{nourbakhsh2020}. 
 
\subsection{Identification of tabular structure}
\begin{figure}[]
  \centering
  \includegraphics[scale=0.4]{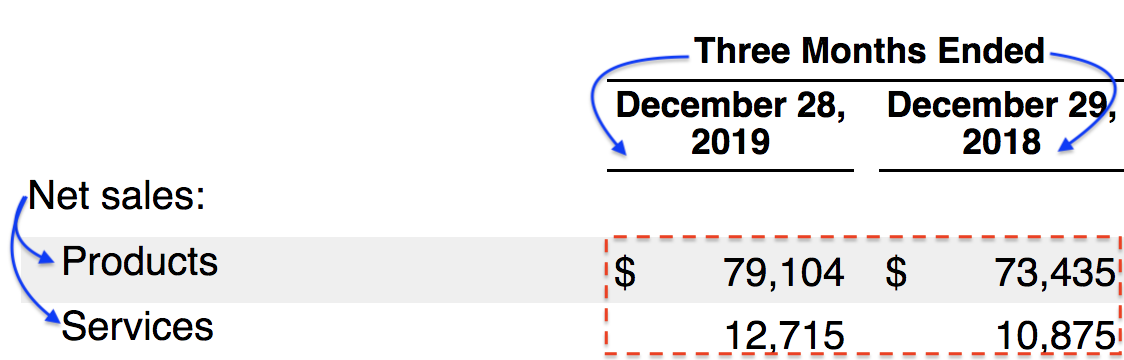}
  \caption{Typical structure of a financial table. The dashed red box identifies the body of the table. The remaining cells are row- and column-headers. The solid blue arrows illustrate the hierarchy of headers. For instance the left-indentation indicates that ``Products'' falls under ``Net sales''. The merged column header indicates that ``Three Months Ended'' applies to both sub-headers below.}
\label{fig:structure}
\end{figure}

It is important for the system to understand the structure of each table, including the distinction between the body of the table and row/column headers. This is done using a rules-based method that finds the largest rectangle in a table that includes numeric information. That rectangle is treated as the body of the table, and the cells that fall outside of the rectangle are treated as row headers or column headers (see Figure \ref{fig:structure} for an example).

Indentation and spatial information are often used to indicate hierarchy in tables (see Figure \ref{fig:structure} for an example). We use the headless Selenium webdriver\footnote{\url{https://www.selenium.dev/selenium/docs/api/py/index.html}} with PhantomJS integration\footnote{\url{https://phantomjs.org/}} to render each table in a background process. This allows us to locate the x- and y-coordinates of every cell in the table, which identifies the alignments of each cell against its row and column headers. This information is used to infer the hierarchy of headers. As an example, the second row header in Figure \ref{fig:structure} is normalized to \texttt{``Net sales --> Products''}.

\section{Operating Segment Identification}
We first narrow the pool of tables down to those likely to include operating segments in them. This is done in two steps:
\begin{enumerate}
    \item Tables not including any financial data, currency or periods in them (such as those with names of board members or office locations) are removed from the pool.
    \item Tables with boilerplate language (i.e. those that are unlikely to include any company-specific language in them) are removed from the pool. This is done by following Algorithm  \ref{algo:stage1_score_table}. The inspiration is that by treating each company as a document, TF-IDF weights can be calculated for each term in each table. These weights would indicate how specific the term is to the company. Tables with a higher aggregate TF-IDF weight are likelier to have operating segments in them. 
\end{enumerate}

\subsection{Data}
We collected 225 earnings reports published between May 1, 2016 and May 1, 2019. The reports belonged to 149 publicly traded U.S. companies within 6 sectors, three belonging to consumer-focused industries (Technology, Media, Retail), and three in the commodities space (Oil/Gas, Metals/Mining, Chemicals). The sectors were determined according to S\&P's standard industry classification\footnote{\url{https://www.spglobal.com/ratings/en/sector/corporates/corporate-sector}}. Among these filings, we extracted 3,124 tables in total. Next, we collected 51,937 individual row/column headers from these tables. Four human annotators manually labeled each header as including or not including an operating segment. To avoid data leakage between training and test sets, 30 companies were set aside for testing. No filings from these companies were included in the training set. Table \ref{tab:data} summarizes the stats of the splits.


\subsection{Header Classification}\label{class}

To address the challenges mentioned in Section \ref{intro}, we approached the problem as that of identifying headers that do \textit{not} include operating segments. This would allow us to focus on metrics that are \textit{not} company-specific, i.e. can occur in any financial report. To do this, we trained a recurrent model with bidirectional GRU units \cite{Cho2014} with the parameters identified in Table \ref{tab:params}. During training, we first built a vocabulary using tokens from the non-operating headers. Next, we iterated through the operating headers and compared each token to the vocabulary. Those present in the vocabulary were left unchanged. Those not found in the vocabulary were masked with the ``<UNK>'' token. This allowed the model to distinguish between common financial terms (such as ``revenue'') and those that were likelier to appear in operating segments (such as ``iPhone'').




\subsection{Evaluation}
The model was trained according to parameters listed in Table \ref{tab:params}. Two configurations were tested using pre-trained GloVe \cite{pennington2014glove} 300d embeddings  and pre-trained ELMo \cite{Peters:2018} embeddings . The resulting models were benchmarked against a suite of baselines listed in Table \ref{tab:perform_summary}. To increase the precision of the models, the operating segment class was treated as the negative class, and evaluation was aimed at high recall for the positive class. This would result in high precision for the negative class. As Table \ref{tab:perform_summary} shows, the recurrent model with pre-trained GloVe embeddings outperformed all baselines in both precision and recall. 

Table \ref{tab:perform_detail} shows a detailed view of the model's performance per sector. F1 performance is relatively consistent overall, with consumer industries doing slightly better than commodities. This might be associated with the fact that operating segments are a lot less company-specific in the commodity market (e.g. ``natural gas'') than in the consumer market (e.g. ``iPhone''). 




\section{USER INTERFACE}
Figure \ref{fig:ui} illustrates how the system would filter operating segments and display them to the end user. Users are presented with a split-screen, where the left-hand panel displays data associated with the operating segments and the right-hand panel connects the data to the earnings reports where it was generated. Users have the ability to review, adjust, and export the data for their analytical purposes. 

\section{CONCLUSION AND FUTURE WORK}
In this paper, we presented SPot, a tool for extracting operating segments from earnings reports in real-time. The tool allows us to trace and record company performance at a granular level. We hope to further enhance SPot's capabilities by normalizing the operating segments into an ontological structure. The insights extracted by SPot can be used for predicting the future performance of a company, identifying potential competitors in the market, and analyzing sector-level trends.

\bibliographystyle{ACM-Reference-Format}
\bibliography{draft}


\begin{thebibliography}{4}


\ifx \showCODEN    \undefined \def \showCODEN     #1{\unskip}     \fi
\ifx \showDOI      \undefined \def \showDOI       #1{#1}\fi
\ifx \showISBNx    \undefined \def \showISBNx     #1{\unskip}     \fi
\ifx \showISBNxiii \undefined \def \showISBNxiii  #1{\unskip}     \fi
\ifx \showISSN     \undefined \def \showISSN      #1{\unskip}     \fi
\ifx \showLCCN     \undefined \def \showLCCN      #1{\unskip}     \fi
\ifx \shownote     \undefined \def \shownote      #1{#1}          \fi
\ifx \showarticletitle \undefined \def \showarticletitle #1{#1}   \fi
\ifx \showURL      \undefined \def \showURL       {\relax}        \fi
\providecommand\bibfield[2]{#2}
\providecommand\bibinfo[2]{#2}
\providecommand\natexlab[1]{#1}
\providecommand\showeprint[2][]{arXiv:#2}

\bibitem[\protect\citeauthoryear{Cho, van Merrienboer, Gülçehre, Bahdanau,
  Bougares, Schwenk, and Bengio}{Cho et~al\mbox{.}}{2014}]%
        {Cho2014}
\bibfield{author}{\bibinfo{person}{Kyunghyun Cho}, \bibinfo{person}{Bart van
  Merrienboer}, \bibinfo{person}{Çaglar Gülçehre}, \bibinfo{person}{Dzmitry
  Bahdanau}, \bibinfo{person}{Fethi Bougares}, \bibinfo{person}{Holger
  Schwenk}, {and} \bibinfo{person}{Yoshua Bengio}.}
  \bibinfo{year}{2014}\natexlab{}.
\newblock \showarticletitle{Learning Phrase Representations using RNN
  Encoder-Decoder for Statistical Machine Translation}. In
  \bibinfo{booktitle}{\emph{Proceedings of EMNLP}}.
  \bibinfo{pages}{1724--1734}.
\newblock


\bibitem[\protect\citeauthoryear{Nourbakhsh, Ghassemi, and
  Pomerville}{Nourbakhsh et~al\mbox{.}}{2020}]%
        {nourbakhsh2020}
\bibfield{author}{\bibinfo{person}{Armineh Nourbakhsh},
  \bibinfo{person}{Mohammad~M. Ghassemi}, {and} \bibinfo{person}{Steven
  Pomerville}.} \bibinfo{year}{2020}\natexlab{}.
\newblock \showarticletitle{SPread: Automated Financial Metric Extraction and
  Spreading Tool from Earnings Reports}. In
  \bibinfo{booktitle}{\emph{Proceedings of the 13th International Conference on
  Web Search and Data Mining}}. \bibinfo{pages}{853–856}.
\newblock


\bibitem[\protect\citeauthoryear{Pennington, Socher, and Manning}{Pennington
  et~al\mbox{.}}{2014}]%
        {pennington2014glove}
\bibfield{author}{\bibinfo{person}{Jeffrey Pennington},
  \bibinfo{person}{Richard Socher}, {and} \bibinfo{person}{Christopher~D.
  Manning}.} \bibinfo{year}{2014}\natexlab{}.
\newblock \showarticletitle{GloVe: Global Vectors for Word Representation}. In
  \bibinfo{booktitle}{\emph{Proceedings of EMNLP}}.
  \bibinfo{pages}{1532--1543}.
\newblock


\bibitem[\protect\citeauthoryear{Peters, Neumann, Iyyer, Gardner, Clark, Lee,
  and Zettlemoyer}{Peters et~al\mbox{.}}{2018}]%
        {Peters:2018}
\bibfield{author}{\bibinfo{person}{Matthew~E. Peters}, \bibinfo{person}{Mark
  Neumann}, \bibinfo{person}{Mohit Iyyer}, \bibinfo{person}{Matt Gardner},
  \bibinfo{person}{Christopher Clark}, \bibinfo{person}{Kenton Lee}, {and}
  \bibinfo{person}{Luke Zettlemoyer}.} \bibinfo{year}{2018}\natexlab{}.
\newblock \showarticletitle{Deep contextualized word representations}. In
  \bibinfo{booktitle}{\emph{Proceedings of NAACL-HLT}}.
  \bibinfo{pages}{2227–2237}.
\newblock


\end{thebibliography}

\begin{algorithm}[]
 \caption{ \label{algo:stage1_score_table} Scoring and filtering of tables based on how likely they are to contain operating segments. The threshold $\delta$ is tuned on validation data for high recall.}
\SetAlgoLined
 initialization\;
 \state $C \gets$ set of all companies\;
 \state $V \gets $ vocabulary\;
 \state $\underset{|V|\times |C|}{\mathrm{W}} \gets$ empty matrix\;
 \For{$v \in V$}{
    \For{$c \in C$}
    {$d_c \gets$ a doc created by merging all filings for $c$\;
    $W_{v,c} = tfidf(v,d_c)$\;}
    }
table filtering\;
 \For{$c \in C$}{
    \For{$f \in $ filings from $c$}{
        \For {$t \in $ tables in $f$}{
            \state $s \gets$ empty array\;
            \For {$r \gets$ sliding window of 2 rows in $t$}{
                \state $text_r = $ set of words in $r$\;
                $s_r = \sum\limits_{v \in text_r} W_{v,c}$\;
            }
        }
        $s_{max} =$ max$(s)$\;
        \eIf{$s_{max}>\delta$}{emit $t$\;}{ignore $t$\;}
    }
 } 
\end{algorithm}

\begin{table}[]
\begin{tabular}{cccc}
\hline
               & \textbf{\# Companies} & \textbf{\begin{tabular}[t]{@{}c@{}}\# Headers with\\op segs\end{tabular}} & \textbf{\begin{tabular}[t]{@{}c@{}}\# Headers\\ without\\ op segs\end{tabular}} \\ \hline
\textbf{Train} & 119                   & 8,333                                                                                 & 33,939                                                                                   \\
\textbf{Test}  & 30                    & 1,618                                                                                 & 8,047                                                                                    \\ \hline
\textbf{Total} & 149                   & 9,951                                                                                 & 41,986                                                                                  
\end{tabular}
\caption{Train/test split in training dataset.}
\label{tab:data}
\end{table}

\begin{table}[]
\begin{tabular}{ccccc}
\toprule 
 & Precision & Recall & F1 & F1(Micro)\tabularnewline
\midrule 
TF-IDF & 0.798 & 0.587 & 0.676 & 0.551\tabularnewline
Random Forest & 0.916 & 0.811 & 0.860 & 0.789\tabularnewline
Logistic Regression & 0.935 & 0.874 & 0.904 & 0.851\tabularnewline
NBC & 0.915 & 0.928 & 0.922 & 0.874\tabularnewline
XGBoost & 0.910 & 0.952 & 0.930 & 0.886\tabularnewline
RNN (ELMo) & 0.969 & 0.979 & 0.974 & 0.958\tabularnewline
RNN (GloVe) & \textbf{0.981} & \textbf{0.983} & \textbf{0.982} & \textbf{0.971}\tabularnewline
\bottomrule
\end{tabular}
\caption{Performance Summary. The F1 column shows the F1 score with the operating segment class being treated as the negative class. The F1(Micro) column shows the F1 score, micro-averaged across the classes.}
\label{tab:perform_summary}
\end{table}


\begin{table*}[]
\begin{tabular}{ll|ll}
\textbf{Param}      & \textbf{Value} & \textbf{Param}         & \textbf{Value}                      \\ \hline
Sequence length     & 25             & Dropout \& pooling     & 0.2 \& concatenated average and max \\
Embedding size      & 300            & Dense layer activation & swish                               \\
No. of hidden units & 50             & Optimizer              & Adam with initial LR of 0.001       \\
Batch normalization & True           & No. of epochs          & 30 (early stopping at 7)           
\end{tabular}
\caption{Final model configuration and training parameters.}
\label{tab:params}
\end{table*}

\begin{table*}[htp]
\begin{tabular}{lll} \\ \hline
\textbf{Example}    & \textbf{Notes on potential challenges}                                                                 & \\ \hline
``iPhone 7''            & Is company-specific; might be renamed to or bundled with iPhone 8 or iPhone X in a more recent report. &  \\
``Natural gas''         & Is not a company-specific segment; other companies might have the same segment.                        &  \\
``Cloud revenue''            & Includes an operating segment (cloud service) and an associated performance metric (revenue). &  \\
``China''               & Is not necessarily an operating segment. Could be referring to the location of an asset or office.     &  \\
``Market intelligence'' & Not necessarily a named entity. Can refer to the concept of business intelligence, or to a segment.             & \\ \hline
\end{tabular}
\caption{A few example operating segments that characteristics that make it difficult to create a generalizable model for them.}
\label{tab:examples}
\end{table*}

\begin{table*}[htp]
\begin{tabular}{c|cccc|cccc}
\toprule 
 & \multicolumn{4}{c}{Commodities} & \multicolumn{4}{c}{Consumer}\tabularnewline
\cmidrule{2-9} \cmidrule{3-9} \cmidrule{4-9} \cmidrule{5-9} \cmidrule{6-9} \cmidrule{7-9} \cmidrule{8-9} \cmidrule{9-9} 
 & Metal & Chemicals & Oil\&Gas & Sector F1 & Tech & Media & Retail & Sector F1\tabularnewline
\midrule 
TF-IDF & 0.816 & 0.760 & 0.792 & 0.789 & 0.601 & 0.692 & 0.589 & 0.611\tabularnewline
Random Forest & 0.776 & 0.831 & 0.810 & 0.809 & 0.891 & 0.808 & 0.920 & 0.883\tabularnewline
Logistic Regression & 0.862 & 0.924 & 0.864 & 0.878 & 0.924 & 0.836 & 0.961 & 0.916\tabularnewline
NBC & 0.918 & 0.913 & 0.892 & 0.902 & 0.942 & 0.831 & 0.971 & 0.931\tabularnewline
XGBoost & 0.932 & 0.901 & 0.931 & 0.924 & 0.944 & 0.845 & 0.959 & 0.933\tabularnewline
RNN (ELMo) & \textbf{0.989} & 0.922 & 0.965 & 0.959 & 0.981 & 0.973 & \textbf{0.997} & 0.981\tabularnewline
RNN (GloVe) & 0.980 & \textbf{0.956} & \textbf{0.983} & \textbf{0.977} & \textbf{0.985} & \textbf{0.985} & \textbf{0.997} & \textbf{0.986}\tabularnewline
\bottomrule
\end{tabular}
\caption{Performance by Sectors (F1 calculated based on operating segments being considered the negative class)}
\label{tab:perform_detail}
\end{table*}

\end{document}